# Spherically Symmetric de Sitter Solution of Black Holes

M F  Mourad[a,b],  N H  Hussein[c], D A Eisa[d], and T A S Ibrahim[a]

.

(a):Mathematics Department, Faculty of Science, Minia University, Minia, Egypt.

(b):mohamed.fouad@mu.edu.eg

(c):Mathematics Department, Faculty of Science, Assiut University, Egypt.

(d):Mathematics Department, Faculty of Science, Assiut University, New Valley, Egypt.

## Abstract

In this study we obtain the solution of the spherically symmetric de Sitter solution of black holes using a general form of distribution functions which include Gaussian, Rayleigh, and Maxwell-Boltzmann distribution as a special case. We investigate the properties of thermodynamics variables such as the Hawking temperature, the entropy, the mass and the heat capacity of black holes. Moreover, we show that the strong energy condition which includes the null energy condition is satisfied. Finally, we show the regularity of the solution by calculating the scalar curvature and invariant curvature in general distribution form.

**Pacs Numbers:** 04.06.kz; 04.06'-m; 04.07.Dy.

## 1. Introduction

One of the greatest contributions of general relativity was the discovery that black holes (BHs) have of thermodynamics (TD) properties. Moreover, for a quantum Schwarzschild black the entropy expressed in terms of the effective mass [1]. It is found that if temperature of black hole (BH) is larger than a critical value, Anti-de Sitter (AdS) (BH) occurs while a thermal gas solution happens when its temperature is less than this value. A (BH) has a natural temperature related to its surface gravity and the entropy associated with its area is non-decreasing.

Hawking and Page formulated the first order phase transition which takes place between the (BH) solution and the thermal gas solution in the (AdS) [2]. It's really helpful for us to understand the holographic description of asymptotically (AdS) space-times [3]. In two and four dimensions there are many solutions for a (BH). The researchers believe that in three dimensions of space-time, these solutions do not exist. However, Deser et. al. explained in the ref. [4], it was possible to find a vacuum solution of Einstein's gravity that could be considered a (BH). The solution is characterized by a constant curvature, but the global topology is different from that of the 3D (AdS) solution. Thus, the causal structure of the solution is closer to Schwarzschild solution [5].

Actually, in Gaussian distribution there exist a new inner horizon in the (BH) space-time, but anisotropic (i.e. $p_r \neq p_\theta$) smearing point of matter with energy momentum tensor (EMT) of anisotropic fluid $T_{\alpha\beta} = diag\ (\rho, -p_r, -p_\theta, -p_\varphi)$, in a "self-consistent" way [6 - 7]. Park [8] investigated the de sitter (dS) in three dimensions $dS_3$ (BHs) for Non Gaussian smearing of hairs and gravastars for Gaussian distribution. In the ref. [9], the deformed $dS_3$ (BHs) have been established which depend on a Rayleigh distribution. Rahman and et al. [10] discussed the effects of the electromagnetic field on the event horizon location, and the properties of mass and (TD) of $dS_3$ (BHs).

The geometric and (TD) properties of the solutions of third-order Lovelock-AdS (BHs) were discussed by Hendi and Dehghani [11]. They noticed that the preserved quantities and (TD) of their (BH) solutions achieved the first law of (TD).

In ref. [12], the proof of the Hawking radiation spectrum cannot be demonstrated by a distribution Bose-Einstein but by the Maxwell-Boltzmann distribution. Jin Bo et al. [13] proved that in the non-static symmetrically background, Hawking temperature correction and tunnel rate correlation is closely connected with the angle parameters of the (BH) background horizon. In recent years [14]-[23], several solutions and their (TD) properties of AdS (BH) have been investigated.

The paper is organized as follows, in section 2, we obtain the solution of the spherically symmetric dS of (BH) using a general form of distribution functions. The (TD) properties of the generalized distribution are discussed in section 3. In section 4, the energy condition of generalized distribution are studied. Finally conclusions of the results are explained in section 5.

## 2. Equations of motion in de Sitter space

The equations of motion with positive cosmological constant are given by Einstein's field equations in the form [24]

$$R_{\alpha\beta} - \tfrac{1}{2}g_{\alpha\beta}R + \Lambda g_{\alpha\beta} = 8\pi G T_{\alpha\beta} \qquad (2.1)$$

where $R_{\alpha\beta}$ ($\alpha,\beta = 0,1,2,3$) is the Ricci tensor, $g_{\alpha\beta}$ is the metric tensor, $R$ is the scalar, $T_{\alpha\beta}$ is the (EMT), $\Lambda = \frac{1}{l^2} = -8\pi P_{th}$ is the positive cosmological constant, $l$ is the radius curvature and $P_{th}$ is the thermodynamic pressure. In order to solve equations (2.1), we consider the metric of spherically symmetric dS space-time of (BHs) in the form

$$ds^2 = f(r)dt^2 - \frac{1}{f(r)}dr^2 - r^2 d\theta^2 - r^2 \sin^2\theta\, d\varphi^2 \qquad (2.2)$$

We noted that $g_{00} = -g_{11}^{-1}$, but this may not true for arbitrary $T_{\alpha\beta}$. So, we consider the specific matter configuration is anisotropic fluid which does not deform (2.2). The (EMT) can be written in the form

$$T_{\alpha\beta} = diag(\rho f, -P_r f^{-1}, -P_\theta r^2, -P_\varphi r^2 \sin^2\theta). \qquad (2.3)$$

Where $\rho$, $P_r$, $P_\theta$ and $P_\varphi$ are functions of $r$. Using the equation (2.2), we obtain the components of Ricci tensor,

$$R_{00} = -f\left(\frac{f''}{2} + \frac{f'}{r}\right)$$

$$R_{11} = f^{-1}\left(\frac{f''}{2} + \frac{f'}{r}\right)$$

$$R_{22} = f'r + f - 1$$

$$R_{33} = \sin^2\theta(f'r + f - 1) \qquad (2.4)$$

and scalar is

$$R = -f'' - \frac{4}{r}f' - \frac{2}{r^2}f + \frac{2}{r^2}, \tag{2.5}$$

where prime (') denotes the derivative with respect to $r$. Substituting from equations (2.2) - (2.5) into (2.1), we get the equations of motion

$$\frac{f'}{r} + \frac{f}{r^2} - \frac{1}{r^2} = -\frac{1}{l^2} + 8\pi G\rho \tag{2.6}$$

$$\frac{f'}{r} + \frac{f}{r^2} - \frac{1}{r^2} = -\frac{1}{l^2} - 8\pi G P_r \tag{2.7}$$

$$\frac{f''}{2} + \frac{f'}{r} = -\frac{1}{l^2} - 8\pi G P_\theta \tag{2.8}$$

$$\frac{f''}{2} + \frac{f'}{r} = -\frac{1}{l^2} - 8\pi G \tag{2.9}$$

We mention that a sufficient condition to deal with some non-singular (BH) is that there exists a regular center, namely, for r → 0, $f(r) = 1 + ar^n$, with n ≥ 2.

The solutions of $P_r, P_\theta, P_\varphi, f(r)$ and $\rho$ can be obtained from (2.6), (2.7), (2.8) and (2.9), namely,

$$P_r = -\rho, \tag{2.10}$$

$$P_\theta = P_\varphi = -\rho - \frac{r}{2}\rho'. \tag{2.11}$$

The general solution of the equation (2.6) is given by

$$f(r) = 1 - \frac{r^2}{3l^2} + \frac{8\pi G}{r}\int_0^r \rho r^2 \, dr \tag{2.12}$$

Now, let we consider the matter density $\rho$ is given by the general distribution function,

$$\rho = A\frac{r^n}{L^{n+3}} e^{-\frac{r^2}{L^2}}, \tag{2.13}$$

where L denotes the characteristic length scale of the material distribution and A denotes the normalization constant.. It is easy to note that for $n = 0, 1, 2$ we obtain Gaussian distribution, Rayleigh distribution, and Maxwell-Boltzmann distribution, respectively. Substituting from (2.13) into (2.12) and integrate, we get

$$f(r) = 1 - \frac{r^2}{3l^2} + \frac{4\pi GA}{r}\gamma\left(\frac{n}{2} + \frac{3}{2}, \frac{r^2}{L^2}\right), \tag{2.14}$$

where

$$\gamma\left(\frac{n}{2}+1, x^2\right) = \int_0^{x^2} t^{\frac{n}{2}} e^{-t} dt$$

and

$$\Gamma\left(\frac{n}{2}+1, x^2\right) = \int_{x^2}^{\infty} t^{\frac{n}{2}} e^{-t} dt = \Gamma\left(\frac{n}{2}+1\right) - \gamma\left(\frac{n}{2}+1, x^2\right)$$

are called the incomplete lower and upper Gamma function respectively.

Comparing the ground state of our solution (equation (2.14)) with the vacuum solution (Schwarzschild solution) in four dimensions $dS_4$ space-time which is given by

$$f(r) = 1 - \frac{r^2}{3l^2} - \frac{2MG}{r}, \qquad (2.15)$$

where $M$ is the total mass of the (BH). The calculation in the background of the vacuum solution ($L \to 0$) is

$$f(r) = 1 - \frac{r^2}{3l^2} + \frac{4\pi G A}{r} \Gamma\left(\frac{n}{2}+\frac{3}{2}\right), \qquad (2.16)$$

we obtain the normalization constant in the following form

$$A = -\frac{M}{2\pi \Gamma\left(\frac{n}{2}+\frac{3}{2}\right)}. \qquad (2.17)$$

Thus, we can write (2.14) as

$$f(r) = 1 - \frac{r^2}{3l^2} - \frac{2MG}{r}\left[1 - \frac{\Gamma\left(\frac{n}{2}+\frac{3}{2}, \frac{r^2}{L^2}\right)}{\Gamma\left(\frac{n}{2}+\frac{3}{2}\right)}\right]. \qquad (2.18)$$

with a positive cosmological constant. We can easily notice that this equation reduces to the $dS_4$ metric from the fact that the last term in (2.18) vanishes at $L \to 0$.

The (BH) horizon $r_-$ and the cosmological horizon $r_+$ are located at $f(r) = 0$. So, they must be satisfied, if exist

$$1 - \frac{r_\pm^2}{3l^2} - \frac{2MG}{r}\left[1 - \frac{\Gamma\left(\frac{n}{2}+\frac{3}{2}, \frac{r_\pm^2}{L^2}\right)}{\Gamma\left(\frac{n}{2}+\frac{3}{2}\right)}\right] = 0 \qquad (2.19)$$

For investigating the physical properties of (BHs) $dS_4$, it is necessary to know the detailed formula of the horizon $r_-$. However, it is difficult to solve equation (2.19) from an analytical point of view, and therefore we consider the approximate solution to be a reasonable solution.

# 3. The thermodynamics properties of generalized distribution

In fact, the surface gravity can be defined for a (BH) which event horizon is a Killing horizon. It retains matter at the horizon [25, 26]. In the case of stationary (BHs), it is proportional to the temperature of the Hawking radiation [27].

The surface gravity κ is given by

$$\kappa = \left|\frac{\partial f}{2\partial r}\right| = \left|\frac{r}{3l^2} + \frac{MG}{\Gamma\left(\frac{n}{2}+\frac{3}{2}\right)}\left[\frac{\frac{2}{L}\left(\frac{r}{L}\right)^{n+2}e^{-\frac{r^2}{L^2}}}{r} - \frac{\gamma}{r^2}\right]\right|, \quad (3.1)$$

Thus, the Hawking temperature in dS space is given by

$$T_H = \frac{\hbar\kappa}{2\pi}\big|_{r_\pm} = \frac{Lx_\pm}{6\pi l^2}\left|1 + \left(3\frac{l^2}{L^2} - x_\pm^2\right)\left(\frac{x_\pm^{n+1}e^{-x_\pm^2}}{\gamma\left(\frac{n}{2}+\frac{3}{2},x_\pm^2\right)} - \frac{1}{2x_\pm^2}\right)\right|, \quad (3.2)$$

where $x_\pm = \frac{r_\pm}{L}$.

This temperature is reduced to ordinary dS Schwarzschild temperature $T = \frac{l^2 - r_\pm^2}{4\pi l^2 r_\pm}$, at $L \to 0$.

The systems are in thermal equilibrium with zero temperature at the Nariai limit (where the (BH) horizon and the cosmological horizon are intersect i.e. ($r_- = r_+ \equiv r_{Nl}$)). In order to study the (TD) of (BHs), we must consider the small (BH) case, expanding equation (3.2) for small $r_-$, we get

$$T_H^- = \frac{1}{4\pi r_-}\left\{(n+2) - \left[\frac{nL^2}{3l^2} + \frac{2(n+3)}{(n+5)}\right]\left(\frac{r_-}{L}\right)^2 + O\left(\left(\frac{r_-}{L}\right)^4\right)\right\}. \quad (3.3)$$

From this, we can find the Nariai radius for small (BH) horizon $x_N = \frac{r_-}{L}$ (i.e. where the temperature vanishes) in the form

$$x_N = \sqrt{\frac{3l^2(n+5)(n+2)}{n(n+5)L^2 + 6l^2(n+3)}}. \quad (3.4)$$

Though we do not know exact form of $r_\pm$ in terms of M, n. We note that the temperature (3.2) is expressed in terms of $r_\pm$. Figures (1, 2, and 3) represent Hawking's temperature vs. radial coordinate at $n = 0, 1$ $and$ 2, respectively, which illustrate a comparison between the exact temperature equation (3.2) and the approximation temperature equation (3.3) and the vacuum solution in four dimensions space. Whereas in Figures (4, 5), we plotted the exact temperature and approximate temperature respectively which represent the Gaussian distribution at $n = 0$, Rayleigh distribution at $n = 1$ and the Maxwell-Boltzman distribution at $n = 2$. The exact temperature is reduced to dS Schwarzschild temperature $T = \frac{l^2 - r_\pm^2}{4\pi l^2 r_\pm}$.

From the drawing it is clear that the left curves represent the temperature of the (BH) horizon, and the curves of the right side represent the temperature of the cosmic horizon $r_+$. The systems are in thermal equilibrium with zero temperature at the Nariai limit.

We know that the entropy is proportional to the area of the event horizon [28], the entropy of the (BH) is

$$S \equiv \alpha \frac{4\pi r_-^2}{4\hbar G} \tag{3.5}$$

Where the coefficient is not fixed to the correct one such as the Bekenstien-Hawking entropy of the large (BH) [29], likewise a small (BH) in a higher curvature [30].

From the first law of (TD)

$$dM = T_H dS = \alpha \frac{2\pi r_-}{\hbar G} T_H dr_- \tag{3.6}$$

So,

$$M(r_-) = \int_0^{r_-} \alpha \frac{2\pi r_-}{\hbar G} T_H dr_-$$

$$= \alpha \frac{L^2}{3l^2 G} \int_0^{\frac{r_-}{L}} x^2 \left| L + (3l^2 - x^2 L^2) \left( \frac{x^{n+1} e^{-x^2}}{\gamma\left(\frac{n}{2} + \frac{3}{2}, x^2\right) L} - \frac{1}{2x^2 L} \right) \right| dx. \tag{3.7}$$

For the results to correspond to the vacuum solution, we put $M(0) = 0$, without (BH), i.e. $r_- = 0$. Therefore, as an approximation to the small (BH), one can obtain

$$M(r_-) = \frac{\alpha r_-}{2G} \left\{ n + 2 - \frac{1}{3} \left[ \frac{nL^2}{3l^2} + \frac{2(n+3)}{(n+5)} \right] \left(\frac{r_-}{L}\right)^2 + O\left(\left(\frac{r_-}{L}\right)^4\right) \right\}. \tag{3.8}$$

The heat capacity $C_p = \frac{dM}{dT_H}$ is given by

$$C_p = \frac{2\alpha\pi L^2}{\hbar G} T_H x \frac{dx}{dT_H} = \frac{2\alpha\pi L^2}{\hbar G} T_H x \left(\frac{dT_H}{dx}\right)^{-1}, \tag{3.9}$$

$\frac{dT_H}{dx} = \frac{1}{6L\gamma^2 l^2 x^2} \left| \{3(x^2 L^2 + l^2)\gamma^2 + [(6n+12)l^2 - (2n+8)L^2 x^2 - 12l^2 x^2 + 4L^2 x^4] x^{n+3} \gamma e^{-x^2} + (4L^2 x^2 - 12l^2) x^{2n+6} e^{-2x^2} \} \right|$ (3.10)

and for small (BH) approximation

$$\frac{dT_H}{dx} = -\frac{\hbar}{4\pi L}\left\{\frac{n+2}{x^2} + \left[\frac{nL^2}{3l^2} + \frac{2(n+3)}{(n+5)}\right] + O((x)^4)\right\} \quad (3.11)$$

## 4. Energy conditions for general distribution

Energy conditions do not generally represent physical constraints, but from the point of view of mathematicians, boundary conditions consider that "energy must be positive" [31]. Energy conditions in general are not physically compatible, for example, the observed effects of dark energy are breaking in a strong energy conditions.

In general relativity, energy conditions are used to demonstrate many important theories of (BH), such as the laws of (TD) of a (BH) [32]. A regular (BH) can also be described by energy conditions [31, 33] which the EMT must achieve the corresponding energy.

In the case of regular (BH), many researchers have used three energy conditions, namely the strong energy condition SEC, the dominant energy condition DEC and weak energy conditions WEC [23].

We note that the weak energy condition is satisfied if and only if [9]

$$\rho \geq 0.$$

Also, from (2.13) and the normalization constant we get

$$\rho = -\frac{Mr^n}{2\pi\Gamma\left(\frac{n}{2}+\frac{3}{2}\right)L^{n+3}} e^{-\frac{r^2}{L^2}}.$$

The weak energy condition does not satisfy in the whole space, due to the matter energy density negative. However, less restrictive but more important conditions in the (BH) dynamics, like the strong or null energy condition can be satisfied.

To see this calculate

$$\rho + P_\varphi = -\frac{r}{2}\rho'$$

$$= \frac{Mr^{n-1}[(\frac{2r^2}{L^2} - n)]}{2\pi\Gamma\left(\frac{n}{2}+\frac{3}{2}\right)L^{n+3}} e^{-\frac{r^2}{L^2}} \quad (4.2)$$

$$\rho + \sum_k P_k = 2P_\varphi = \frac{Mr^n[(n+2)-\frac{2r^2}{L^2}]}{2\pi\Gamma\left(\frac{n}{2}+\frac{3}{2}\right)L^{n+3}} e^{-\frac{r^2}{L^2}}. \quad (4.3)$$

We have used $\rho + P_r = 0$ from (2.10). We find that $\rho + P_\varphi \geq 0$, for $\frac{r}{L} \leq \sqrt{\frac{n}{2}}$,

whereas, $\rho + \sum_k P_k \geq 0$, when $\frac{r}{L} \leq \sqrt{\frac{n+2}{2}}$. Hence the SEC which include the null energy condition is satisfied when $\frac{r}{L} \leq \sqrt{\frac{n}{2}}$.

In fact, non-singular (BH) model represents a mathematical theory of the (HBs) that avoids some of the theoretical problems in the standard (BH) model such as information loss and the unobservable nature of the event horizon. Hawking and Penrose demonstrated that space-time with (BH) cannot be completed geodesically [34] and it is clear that there is a relationship between the non-singular (BH) and the curvature and curvature of the curve if they are invariant.

Now, we discuss the regularity of the solution by obtaining the scalar curvature and the invariant curvature [34]

$$R = g^{\mu\nu} R_{\mu\nu} = \frac{4}{l^2} + \frac{2M}{\Gamma\left(\frac{n}{2}+\frac{3}{2}\right)} \left[\frac{2n+8}{L^3}\left(\frac{r}{L}\right)^n - \frac{4}{L^3}\left(\frac{r}{L}\right)^{n+2}\right] e^{-\frac{r^2}{L^2}}. \tag{4.4}$$

$$Rs = R^{uv} R_{abuv} = 2C^2 + 2D^2, \tag{4.5}$$

where

$$C = -\frac{1}{l^2} - \frac{2M}{\Gamma\left(\frac{n}{2}+\frac{3}{2}\right)} \left[\frac{n}{L^3}\left(\frac{r}{L}\right)^n - \frac{2}{L^3}\left(\frac{r}{L}\right)^{n+2} + \frac{2}{L^3}\left(\frac{r}{L}\right)^n\right] e^{-\frac{r^2}{L^2}},$$

$$D = -\frac{1}{l^2} - \frac{4MG}{\Gamma\left(\frac{n}{2}+\frac{3}{2}\right)L^3}\left(\frac{r}{L}\right)^n e^{-\frac{r^2}{L^2}},$$

The contravariant Ricci tensor is given by $R^{\mu\nu} = g^{\mu\alpha} g^{\nu\beta} R_{\alpha\beta}$.

From curves (6-11), we show that the scalar curvature, the metric and invariant curvature are non-singular everywhere. This solution is regular everywhere and the (BH) is non-singular.

## 5. Conclusions

In this paper, we discussed the regular (BH) which has no singularity at the origin in dS space-time. We avert this singularity by considering some of distribution functions. We studied the solution of Einstein equations in smearing of point matter. We investigated the regular solution by considering the general distribution function form; including Gaussian distribution at $n = 0$, Rayleigh distribution at $n = 1$ and Maxwell-Boltzmann distribution at $n = 2$.

Recently Hussein et al, [35] discussed (TD) many variables of some types of (BHs). We studied the (TD) properties of a (BH) by using the relation between the entropy and the area. Also we obtained the mass of the small (BH) from the first law of (TD). In the figures (1-5), we plotted the comparison of Hawking temperature in exact form (3.2), approximated formula (3.3) and vacuum dS Schwarzschild solution respectively. The right-hand side curves represent the temperature due to the cosmological horizon and the left-hand side curves represent the temperature due to (BH) horizon and they are intersected at the Nariai radius.

We proved that the WEC are not satisfied but the SEC and the DEC are satisfied. Finally, we show the regularity of the solution by calculating the scalar curvature and invariant curvature in general distribution form. Also, we plotted the scalar curvature and invariant curvature for Gaussian in figures (6 and 7), and for non-Gaussian are plotted in figures (8,-11). We note that they are regular everywhere.

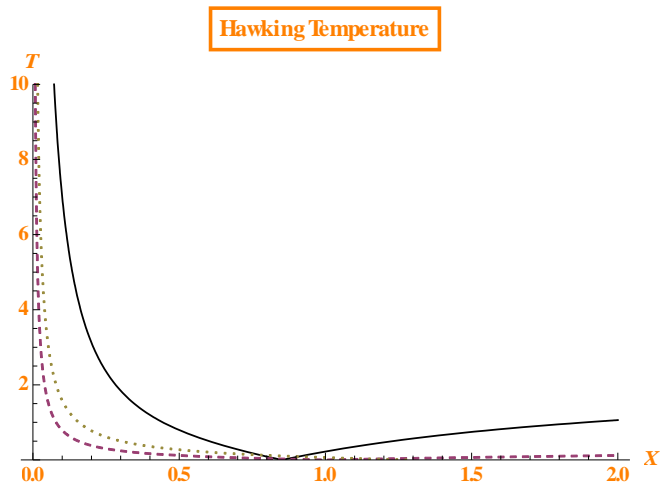

Figure 1: The Hawking temperature for Gaussian distribution ($n = 0$); the solid line denote equation (3.2), the dotted line denote equation (3.3) and dashed line denote the temperature for vacuum in four dimension.

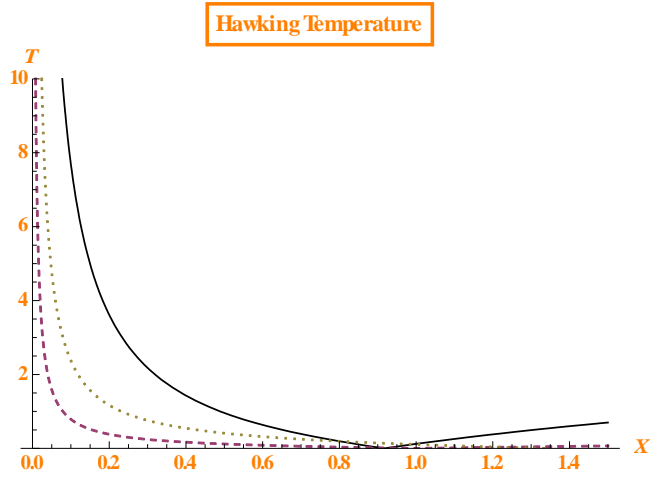

Figure 2: The Hawking temperature for Rayleigh distribution($n = 1$); the solid line denotes equation (3.2), the dotted line denotes equation (3.3) and dashed line denote the temperature for vacuum in four dimension.

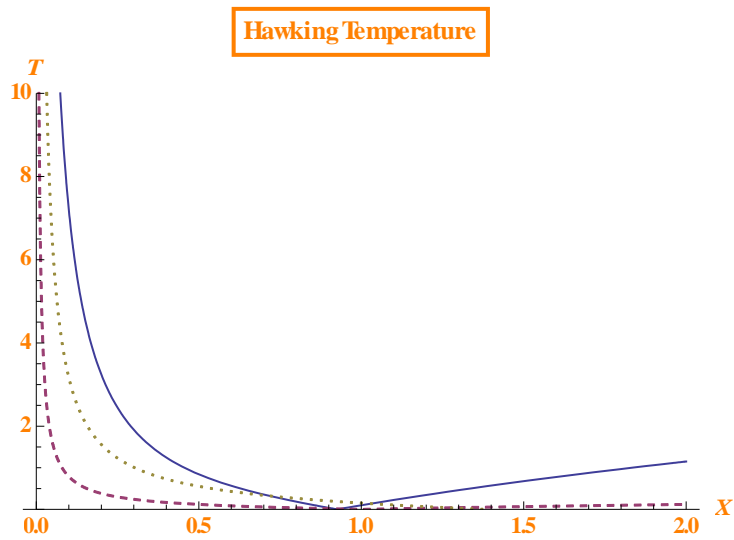

Figure 3: The Hawking temperature for Maxwell-Boltzmann distribution ($n = 2$); the solid line denotes (equation (3.2)), the dashed line denotes (equation (3.3)) and dotted line denotes the temperature for vacuum in four dimension.

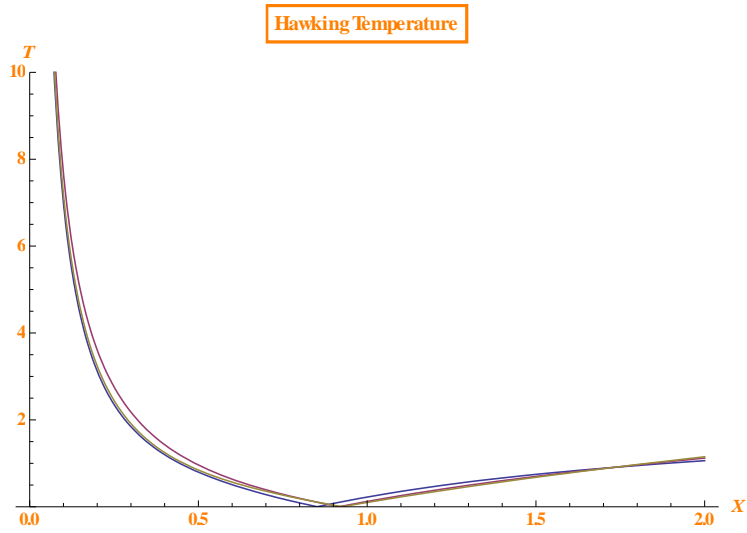

Figure 4: The exact Hawking temperature (equation (3.2)) for $n = 0, 1, 2$ from bottom to top respectively.

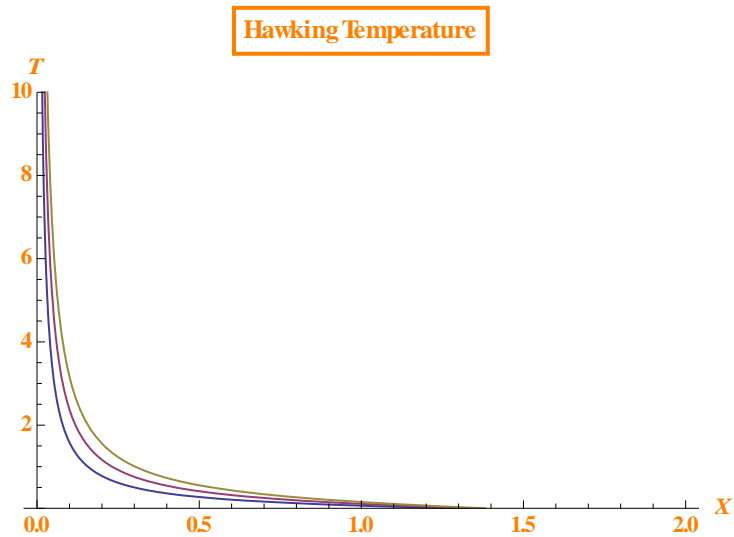

Figure 5: The approximated Hawking temperature (equation (3.3)) for $n = 0, 1, 2$ from bottom to top respectively.

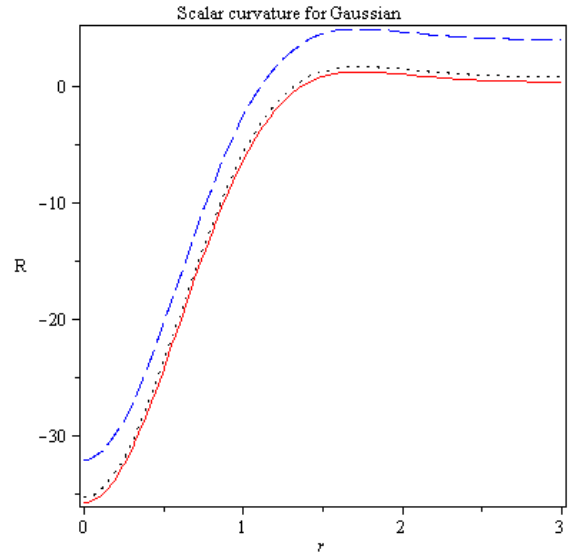

Figure 6: Scalar curvature with radius for Gaussian distribution function under varies values of $l = 1, 5, 10$ which correspond to dashed, dotted and solid line respectively.

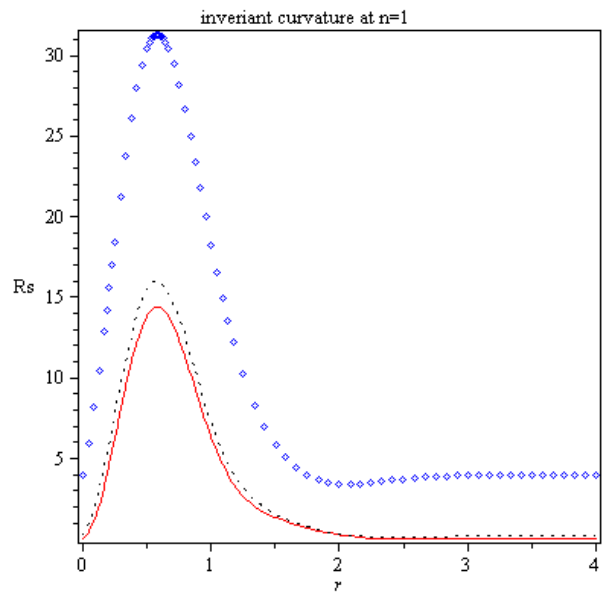

Figure 7: Invariant curvature for Gaussian distribution function under varies values of $l = 1, 5, 10$ presented by solid line, dot line and dashed line respectively.

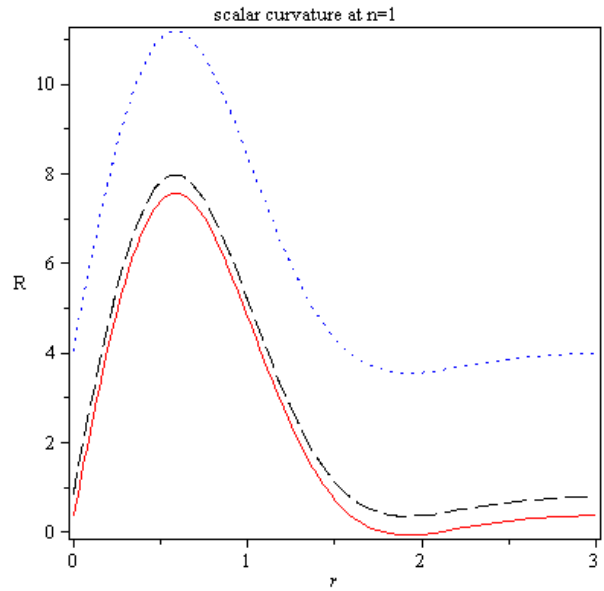

Figure 8: Scalar curvature with radial at $n = 1$ under varies values of $l = 1, 5, 10$ presented by dotted, dashed and solid lines respectively.

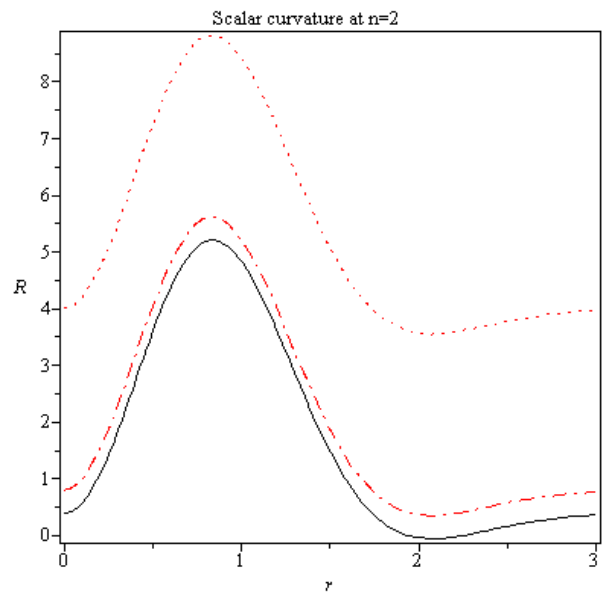

Figure 9: Scalar curvature with radial at $n = 2$ under varies values of $l = 1, 5, 10$ presented by solid, dotted and long dashed lines respectively.

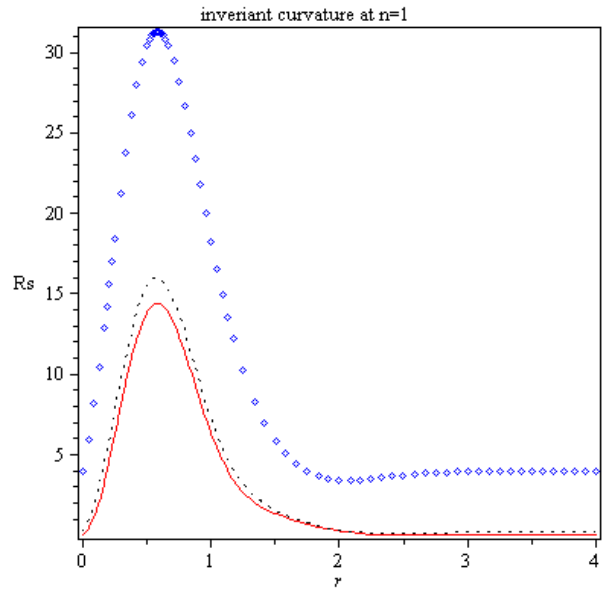

Figure 10: Invariant curvature with radial for $n = 1$ under varies value of $l = 1, 5, 10$ correspond to dotted line, dash line, solid line respectively.

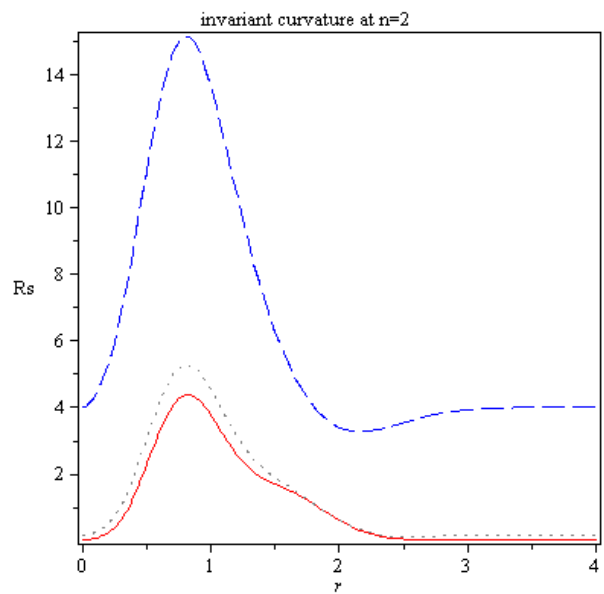

Figure 11: Invariant curvature for $n = 2$ under varies values of $l = 1, 5, 10$ correspond to dashed line, dotted line, solid line respectively.

**Acknowledgements**: We thank Professor N A Hussein, Assiut University in Egypt for the useful discussion and important comments.